\documentstyle[11pt,aaspp4]{article}    

\def\la{\lower.5ex\hbox\frac{$\; \buildrel <}{\sim \;$}}
\def\ga{\lower.5ex\hbox\frac{$\; \buildrel >}{\sim \;$}}
\def\lyal {{\rm Ly} \alpha}
\def\nh {N_{\scriptscriptstyle\rm HI}}
\def\j21{J_{-21}}
\def\tj21{\tilde{J}_{-21}}
\def\etal{et~al.\ }

\begin{document}

\title{ON THE ORIGIN OF METALLICITY IN LYMAN-ALPHA FOREST SYSTEMS}

\author{Masashi Chiba\altaffilmark{1}}
\affil{National Astronomical Observatory, Mitaka, Tokyo 181, Japan}
\altaffiltext{1}{chibams@cc.nao.ac.jp}

\and

\author{Biman B. Nath\altaffilmark{2}}
\affil{Inter-University  Center for Astronomy \& Astrophysics, Post Bag 4,
 Pune 411007, India}
\altaffiltext{2}{biman@iucaa.ernet.in}

\begin{abstract}
We investigate the hypothesis that $\lyal$ absorption lines arise
in two populations of halos --- minihalos of small circular velocity
($V_c \la 55$ km~s$^{-1}$) in which star formation and metal production
are inhibited by photoionization of the UV background radiation,
and large galactic halos ($55 \la V_c \la 250$ km~s$^{-1}$) which possess
stars and metals. Based on the model of $\lyal$-absorbing gas confined in both
populations of halos, we attempt to explain the recent observations of
(1) associations of visible galaxies with $\lyal$ lines
at low redshifts $z \la 1$, and 
(2) metal lines associated with a non-negligible fraction of low H~I column
density $\lyal$ lines at $z \sim 3$.
For galactic halos, we find that photoionized gas clouds confined in the
pressure of ambient hot gas can produce $\lyal$ absorptions with H~I column
density as low as $10^{14}$ cm$^{-2}$, and that the impact parameter of
a sightline for such absorptions matches well with the observed radius of
gaseous envelope in a typical luminous galaxy.
Using the Press-Schechter prescription for the mass function of halos,
we also show that the fraction of $\lyal$ lines with associated metal lines
can be understood in terms of the fraction of $\lyal$ absorbers that are
associated with galactic halos.
In particular, the reported fraction of $\sim 0.5 \hbox{--} 0.75$ at
$z \sim 3$ is reproduced when the boundary value of $V_c$ to separate
mini or galactic halos is $40 \sim 60$ km~s$^{-1}$, which is consistent with
the theoretical prediction of galaxy formation under photoionization.
The average metallicity of both $\lyal$ forest and
damped $\lyal$ systems at $z \sim 3$ is explained in terms of the model
of halo-formation history combined with the age-metallicity relationship of
Galactic halo stars.
Possible methods to test this hypothesis and the other alternative
scenarios are also discussed.
\end{abstract}


\section{INTRODUCTION}

Several recent observations indicate that even the low H~I column density
$\lyal$ absorption lines may not be primordial as were previously thought.
High resolution observations by Cowie \etal (1995), Tytler \etal (1995)
and Womble \etal (1996) show that there are metal lines (C~IV) associated
with $\lyal$ absorption lines with as small H~I column density as
$\nh \sim 10^{14}$ cm$^{-2}$ at $z \sim 3$. The fraction of the $\lyal$ lines
with $\nh \ga 10^{14}$ cm$^{-2}$ that are associated with metal lines
is approximately $\ga 0.5$ at $z \sim 3$.
The median abundance of lines with $\nh \ga 10^{14.5}$ cm$^{-2}$ at
$z \sim 3$, is found to be [C/H] $ \sim -2.5$.
Recently Songaila and Cowie (1996) have shown that the fraction of $\lyal$
lines that are associated with metal lines, hereafter denoted as $f_m$,
could be higher, and they have also detected the other C~II, S~IV, N~V
absorption lines in hydrogen clouds with $N_{HI} > 10^{15}$ cm$^{-2}$.
In view of the dependence of observational sensitivity for detecting
metal lines, the fraction $f_m$ represents only a lower limit for the
fraction of $\lyal$ lines associated with `metals'.

Another set of observations shows that at low redshifts a large fraction of
$\lyal$ absorption lines are associated with galaxies (Maloney 1992;
Morris \etal 1993). Lanzetta \etal (1995b, LBTW) claimed
that $\sim 60 \pm 19 \%$ of the $\lyal$ lines at $z \la 1$
arise in galactic halos. The observation of Barcons \etal (1995)
in which they imaged a galaxy corresponding to an absorption
line, is an evidence of such associations.
It is, therefore, reasonable to assume that at least a fraction of
the $\lyal$ absorption lines arise in galactic halos, and this fraction
is hereafter denoted as $f_g$. 

In the hierarchical model of structure formation, there is a continuum
of halos of various masses and velocity dispersions.
In this paper, we consider the hypothesis that the $\lyal$ lines
are produced in two populations of halos in this continuum -- in `minihalos'
of small velocity dispersions which have not cooled and formed stars, and
in large galactic halos with stars and metals. Based on this two-population
model for QSO absorbers, we attempt to explain $f_m$ in terms of $f_g$
by the relation $f_m \sim f_g$.

There are circumstantial evidences for such a two population
scenario. In the column density distribution of $\lyal$ lines,
there is a break around $\nh \sim 10^{15}$ cm$^{-2}$ (see, e.g.,
Cristiani \etal 1995). Petitjean \etal (1993) have interpreted
this break as being due to existence of two populations of
$\lyal$ lines -- the lines with higher H~I column density
being associated with strong metal lines and with galactic halos.
Recently, Fern\'andez-Soto et al. (1996) analyzed the two-point correlation
of very weak C~IV absorption lines associated with high-redshift ($z \sim 2.6$)
$\lyal$ absorption systems. They found that high-redshift $\lyal$ absorption
systems traced by C~IV lines are clustered in redshift, as strongly as that
expected for galaxies. This suggests that many $\lyal$ absorbers
even at high redshifts may arise in galaxies.

Our primary goal in this paper is therefore to determine whether these recent
observations can be understood within the framework of a two-population model
for $\lyal$ absorption systems, including their redshift evolution in the
context of hierarchical structure formation.
In \S 2, we present a model of $\lyal$-absorbing gas confined in both mini and
galactic halos and set the prescription for determining the number of
absorption lines provided by various masses of halos. On the basis of
the Press-Schechter mass function with various power spectra and cosmological
parameters, we then discuss the fraction of absorption lines in minihalos
and in galactic halos by comparing with the recent Keck observations in \S 3.
The simplified approach for modeling chemical evolution of halos is also
presented to predict the evolution of the average metallicity associated
with $\lyal$ lines in redshift. The discussion and conclusions are finally
drawn in \S 4.

\section{MINIHALOS AND GALACTIC HALOS}
\subsection{{\it Assumptions and definitions}}

We define minihalos as halos in which baryonic gas is entirely photoionized
by the external UV background and the gas is stably confined in the
gravitational field of the dark matter (Rees 1986; Ikeuchi 1986).
In these minihalos, the gas does not go through extensive star formation
because of suppression of radiative cooling by photoionization (Efstathiou
1992). Below, we will characterise the halos by their circular velocities
$V_c$, where the mass density of halos is dominated by extended dark matter
with $r^{-2}$ distribution.
Minihalos are therefore bounded by two scales of circular velocity,
$V_1 < V_c < V_2$. The lower limit is set by the sound
velocity of the photoionized intergalactic medium (IGM), which
is $\sim 15 \, (T_{IGM}/ 10^4 \, {\rm K})^{0.5}$ km~s$^{-1}$.
Halos with $V_c$ less than this value will not have baryonic
gas infall from the IGM. This is basically the Jeans mass, below
which the perturbation in baryonic gas is suppressed.
We take a characteristic value of $V_1=15$ km~s$^{-1}$ for the photoionized
IGM with $T_{IGM} \sim 10^4$ K.

The upper limit $V_2$ is admittedly less certain.
Recent numerical simulations show that photoionization
severely inhibits the formation of galaxies with $V_c \la 30$
km~s$^{-1}$, and substantially decreases the mass of cooled
baryons in halos with $V_c \la 50$ km~s$^{-1}$ (Thoul and
Weinberg 1996). They also find that above $V_c \sim 75 $ km~s$^{-1}$
photoionization of the IGM  does not have any significant effect.
For the reasons described in the next subsection, we will use
a canonical value of $V_2 = 55$ km~s$^{-1}$ and examine
the effect of changing $V_2$ over its possible range.
In fact, we later ask the question whether or not one can {\it
determine} the value of $V_2$ from the observations.
In brief, then, we assume
that halos with $V_c < V_2 \sim 55$ km~s$^{-1}$ do not undergo
vigorous star formation, and call them minihalos in this paper.

We will call the halos with circular velocity larger than $V_2$,
the galactic halos. Our results are not sensitive to the upper limit
of the circular velocity for galactic halos, and we will use a 
value of $V_3=250$ km~s$^{-1}$ as an upper limit. 
Below, we will associate these halos
with $\lyal$ lines which have corresponding metal lines. It is true
that the division of halos into the above categories of mini
and galactic halos is rather ad hoc. However, it is most likely to be
close to the real picture for the above mentioned reasons.

For the gas associated in galactic halos, the virial temperature $T$
for $V_c \ga V_2 \sim 55$ km~s$^{-1}$ is above $10^5$ K [$T \sim 10^5
(V_c/55 kms^{-1})^2$ K], where the mean molecular weight is taken as 0.59
for a totally ionized gas. In this temperature range, any local disturbances
in the hot gas may be thermally unstable, because the radiative cooling rate
increases with decreasing $T$. We therefore assume that the instability
promotes the development of a two-phase structure in the gas (Fall \& Rees
1985), so that it is the cool gas clouds, being pressure-confined by the
surrounding hot gas, that produce $\lyal$ absorptions,
as proposed by Mo (1994).
In this work, we will be concerned with $\lyal$ lines with low HI column
density as $\nh \la 10^{16}$ cm$^{-2}$, in which the shielding against
the UV background radiation is ineffective. The internal density of clouds
in the outer parts of halos that give rise to such lines is low enough to
be photoionized by the UV background radiation. We thus assume the gas in
clouds to be in photoionization equilibrium, where the gaseous temperature
is expected to be order of ${\rm few} \> \times 10^4$ K,
and therefore consistent with the observed Doppler parameters of
$\lyal$ lines (see, e.g., Maloney 1992; Rauch et al. 1996). 

Note that, even the gas in disks, if disks are formed, will be ionized by
the UV radiation in the outer parts, beyond the observed sharp edge
at $\nh \sim {\rm few} \times 10^{19}$ cm$^{-2}$ (Maloney 1992). The
galaxies at low redshift imaged by Barcons \etal (1995) are inferred to 
give rise to absorption lines at an impact parameters of $\ga 50 h^{-1}$ kpc,
where $H_0=100h$ km~s$^{-1}$ Mpc$^{-1}$.
The gas which is responsible for such absorption lines is certainly
ionized by the UV background radiation (see also, Morris \etal 1993;
Salpeter and Hoffmann 1995).

\subsection{{\it H~I column density}}

We assume that the density profile of both mini and galactic halos
is represented by that of a softened isothermal sphere,
\begin{equation}
\rho = \frac{\tilde{\rho} r_c^2}{r^2+r_c^2} \>, 
\end{equation}
where $\tilde{\rho}$ is the central density and $r_c$ is the core radius.
The baryonic gas amounts to a fraction $F$ of the total mass of the halo.
Then, the circular velocity $V_c(r)$ at $r$ can be written as,
\begin{equation}
V_c^2(r) = \frac{GM(r)}{r}=
4\pi G \tilde{\rho} r_c^2 \frac{1}{x} ( x - \arctan x ) \> ,  
\end{equation}
where $x \equiv r/r_c$. This functional form of $V_c(r)$ assures a nearly flat
rotation curve at large radii $r \gg r_c$.
In what follows, the circular velocity is characterised by its value
at the virial radius of the halo, $r_{vir}$ (see below for an explanation
of $r_{vir}$). 
Then, substituting eq.(2) into
eq.(1) and introducing $x_v \equiv r_{vir}/r_c$, the density profile of a halo
that includes both baryonic gas and dark matter is re-written as,
\begin{equation}
\rho(r, V_c) = \frac{V_c^2}{4\pi G (r^2 + r_c^2)}
 \Bigl ( \frac{x_v}{x_v - \arctan x_v} \Bigr ) \> ,  
\end{equation}
where $V_c$ is the circular velocity at $r=r_{vir}$.

We model the gas in minihalos ($V_1 \le V_c \le V_2$) to be in photoionized
equilibrium with the external UV background radiation.
We write the intensity of the UV background radiation at the
Lyman limit as $J=\j21 10^{-21}$ erg cm$^{-2}$ s$^{-1}$ sr$^{-1}$ Hz$^{-1}$.
Using the ionization and recombination rates for a fiducial UV spectrum of
$J_\nu \propto \nu^{-1}$ (Black 1981), the H~I number density $n_{HI}$
in minihalos is written as,
\begin{equation}
n_{HI} = \frac{\alpha_H(T)}{4\pi \j21 10^{-21} G_H}
       \Bigl( \frac{F \rho}{m_p} \Bigr )^2 \qquad {\rm for\ minihalos}\>, 
\end{equation}
where $G_H = 2.54\times 10^8$ for hydrogen and $\alpha_H$ is the recombination
rate coefficient (Black 1981).
The column density of H~I at an impact parameter $b$ is then estimated as
\begin{eqnarray}
N_{HI}(b) &=& 2 \int_{b}^{\infty} \frac{n_{HI} r}{\sqrt{r^2-b^2}}dr \nonumber\\
  &\sim& 3.2 \times 10^{14} \Bigl ( \frac{T}{3\times10^4K} \Bigr )^{-3/4}
\Bigl ( \frac{F}{0.05} \Bigr )^2 J_{-21}^{-1}
\Bigl ( \frac{V_c}{30 km s^{-1}} \Bigr )^4   \nonumber \\
&\times& \Bigl ( \frac{x_v}{x_v-\arctan x_v} \Bigr )^2
 \frac{1}{[(b/10kpc)^2+(r_c/10kpc)^2]^2} \quad {\rm cm}^{-2}
 \quad {\rm for\ minihalos} \> .    
\end{eqnarray}

For galactic halos ($V_2 \le V_c \le V_3$), we assume a cloud confined by
ambient hot gas to be a uniform sphere with proton number
density $n_{cl}$ and temperature $T_{cl}$. The pressure balance with hot gas
with proton number density $n$ ($=F\rho / m_p$) and temperature $T$ yields
$n_{cl} = n T / T_{cl}$. Under the assumption that the clouds are
photoionized by the external UV background and in ionization equilibrium,
the H~I number density $n_{HI}$ is then written as
\begin{equation}
n_{HI} = \frac{\alpha_H(T_{cl}) n_{cl}^2}{4\pi \j21 10^{-21}G_H} 
   \propto \j21^{-1} \alpha_H(T_{cl}) \Bigl ( \frac{T}{T_{cl}} \Bigr )^2 n^2
\qquad {\rm for\ galactic\ halos}\> . 
\end{equation}
The cloud radius $R_{cl}$ for the mass of each cloud $M_{cl}$ is
\begin{equation}
R_{cl} = \Bigl ( \frac{3M_{cl}}{4\pi n_{cl}m_p} \Bigr )^{1/3}
      \propto M_{cl}^{1/3} \Bigl ( \frac{T}{T_{cl}} \Bigr )^{-1/3} n^{-1/3}
\> .  
\end{equation}
Each cloud produces the H~I column density as
$N_{HI}(r) \sim n_{HI}(r)R_{cl}(r)$. In order to obtain the H~I column density
observed at an impact parameter $b$ from the galactic center, it is
required to know the spatial distribution of clouds in halos {\it a priori}.
Without going into details for modeling this, we assume that the clouds have
roughly unit covering factor in halos when projected onto the sky, and that
each sightline encounters only one cloud in average, which is not
in disagreement with the observational results (Morris et al. 1993; LBTW).
Then, as we argue in the Appendix, the average H~I column density observed at
an impact parameter $b$ is well approximated by $N_{HI}(r \to b)$. We thus
obtain from eq.(6) and (7)
\begin{eqnarray}
N_{HI}(b) &\sim& 9.6 \times 10^{15}
 \Bigl ( \frac{M_{cl}}{10^6 M_{\odot}} \Bigr )^{1/3}
 \Bigl ( \frac{T_{cl}}{3\times10^4K} \Bigr )^{-29/12}
 \Bigl ( \frac{F}{0.05} \Bigr )^{5/3} J_{-21}^{-1}
 \Bigl ( \frac{V_c}{200 km s^{-1}} \Bigr )^{20/3}   \nonumber \\
&\times & \Bigl ( \frac{x_v}{x_v-\arctan x_v} \Bigr )^{5/3}
 \frac{1}{[(b/100kpc)^2+(r_c/100kpc)^2]^{5/3}} \quad {\rm cm}^{-2}
 \quad {\rm for\ galactic\ halos} \> ,    
\end{eqnarray}
where we assume the virial temperature for $T$ of the hot gas.
We note that both expressions for $N_{HI}$ given in eq.(5) and (8) are
approximately connected continuously at the canonical value of
$V_c \sim 55$ km~s$^{-1}$ for $V_2$, corresponding to the virial temperature of
$T \sim 10^5$ K; this is the temperature of hot gas below which the picture of
a two-phase structure in the gas becomes invalid. We therefore use eq.(5)
for $V_c < 55$ km~s$^{-1}$ and eq.(8) for $V_c \ge 55$ km~s$^{-1}$
in the following.

According to several determinations of the intensity of the UV background
radiation (e.g. Madau 1992), $\j21$ is nearly constant at high redshifts
$2.5 \la z \la 3.5$, whereas at low redshifts $z \la 2.5$ it rapidly decreases
with decreasing redshift. We adopt the following functional form for $\j21$,
\begin{equation}
\j21 (z) = \left\{
  \begin{array}{ll}
  \tj21 &\qquad\mbox{for $z \geq 2.5$}  \nonumber \\
  \tj21 \Bigl ( \frac{1+z}{3.5} \Bigr )^\alpha 
  &\qquad\mbox{for $z < 2.5$} \>,
  \end{array}\right.
\end{equation}
where the index $\alpha$ quantifies the decreasing rate of $\j21$ for $z<2.5$.

To estimate the value of $x_v (=r_{vir}/r_c)$ in eq.(3),
we use the theory of the spherical collapse of cosmological density
perturbations, that yields $r_{vir}=0.1 H_0^{-1} (1+z)^{-3/2} V_c$ for the
Einstein-de Sitter (EdS) universe (e.g. Peebles 1980; Padmanabhan 1993).
This allows us to
derive $x_v$ using the observed values of $V_c$ and $r_c$ of the
present-day dark matter halos at $z=0$. For this purpose, it is convenient to
use dark-matter dominated dwarf spirals (Moore 1994),
where the dynamical effect of a luminous component on the observed rotation
curve is sufficiently small, so that one can accurately determine the
structural parameters of dark halos. Adopting the values of $V_c$ and $r_c$
from Moore (1994) for dwarf spirals of DDO154, DDO170, DDO105 and NGC3109,
we obtain $10 \la x_v \la 20$ for the EdS universe with $h=0.5$.
In what follows, we assume $x_v=15=const.$, since the following
results in the concerned low-column density range of
$N_{HI} \la 10^{16}$ cm$^{-2}$ are quite insensitive to the values of $x_v$
as long as $x_v \gg 1$.

Figure 1(a) shows the impact parameter $b$ {\it vs.} the total mass of halos
$M$ at $z = 0.5$, where we take $T=3\times10^4$ K for the gas in minihalos
(eq.5), $T_{cl}=3\times10^4$ K and $M_{cl}=10^6 M_{\odot}$ for the clouds in
galactic halos (eq.8), with the common parameters of $F=0.05$, $\tj21=1$,
$\alpha=2$, and $x_v=15$. It is seen that lines of constant $N_{HI}$ have
the slightly different slope at $M \la 2.3\times 10^{10} M_{\odot}$
(or $V_c \la 55$ km~s$^{-1}$) from that at larger $M$ or $V_c$. This is because
$N_{HI} \propto b^{-3}$ in eq.(5) while $\propto b^{-10/3}$ in eq.(8)
for $b \gg r_c$. Here we note that the discontinuities at the boundary of both
regimes (at $V_c \sim 55$ km~s$^{-1}$) are quite small in $b$ so that these
will not affect the later results.  We also find that H~I column
densities as low as $N_{HI} \sim 10^{14}$ cm$^{-2}$ can be provided by either
the low-mass minihalos with the impact parameter $b$ below $\sim 40$ kpc or by
the high-mass galactic halos with larger $b$. This is more clearly presented in
Fig.1(b) that shows $N_{HI}$ {\it vs.} $b$ for a given total mass $M$.
Dotted line denote the observed size of gaseous envelopes
$\sim 160$ $h^{-1}$ kpc in luminous galaxies which give rise to $\lyal$
absorptions at $z \la 1$ (LBTW). In our model of absorbers, this is achieved
for a mass range of $10^{11} \la M \la 10^{12}$ $M_\odot$, which corresponds to
galaxies with luminosity $10^{9} \la L \la 10^{10}$ $L_\odot$ for the
baryon fraction $F$ of 0.05 and the mass-to-light ratio of the order of 5.
Thus, the model yields the right order of the size of extended
$\lyal$-absorbing gaseous envelopes in galaxies.

Bold solid line in Fig.1(b) corresponds to the observed correlation between
the H~I column density and impact parameter by LBTW, assuming Doppler
parameters of absorption lines in the range of $20-40$ km~s$^{-1}$.
The present model also implies the correlation and agrees approximately with
LBTW, where the slight discrepancy in the slope may be due to the effects of
sampling over different galactic masses and/or impact parameters.
Recently Le Brun \etal (1996) and Bowen \etal (1996) claimed
that the equivalent width of $\lyal$ lines do not correlate
with the impact parameter. This has been argued to weaken the possibility of
$\lyal$ lines being directly associated with galaxies. The other possibilities
discussed in the literature include the scenario of filaments between luminous
galaxies, and $\lyal$ lines from outflows from galaxies. The issuse may be
settled by assembling the statistically meaningful number of
$\lyal$ absorber-galaxy pairs and by comparing with the current type model
that presents the explicit dependence of an impact parameter on the masses of
galaxies and redshifts (Fig.1c).

\subsection{{\it Cosmological context}}

With the knowledge of the H~I column density that is produced at an impact
parameter $b(\nh , V_c)$, we can calculate the number of $\lyal$ lines with
a column density larger than $\nh$, if the number density of halos with
circular velocity $V_c$ is known. We use the Press-Schechter formalism for
this purpose, and CDM models of structure formation (similar to the work of
Mo et al. 1993).

For the EdS universe, the circular velocity $V_c$ and mass $M$ of a CDM halo
are related as
\begin{equation}
M= \frac{4\pi}{3} \, \rho_0 r_0 ^3 \, , \qquad
V_c=1.67 \, (1+z)^{0.5} \, H_0 r_0 \>,   
\end{equation}
where $\rho_0$ is the present-day mean density of the universe and $r_0$ is
the comoving radius of the halo. The comoving number density of halos per unit
circular velocity $V_c$ is given by (Mo et al. 1993)
\begin{eqnarray}
n(V_c, z) \, dV_c &=& \frac{-3 (1.67)^3 \, \delta_c \, H_0 \, (1+z)^{5/2}}{
(2 \pi)^{3/2} \, V_c^4 \, \Delta(r_0) } \, \frac{d\ln \Delta}{d\ln V_c}
  \nonumber \\
&\times & \exp \Bigl( \frac{ - \delta _c^2 (1+z)^2}{2 \, \Delta ^2 (r_0)}
\Bigr ) \> dV_c \>. 
\end{eqnarray}
Here $\Delta (r_0)$ and $\delta _c=1.68$ are the rms and threshold linear
overdensities in a spherical region of radius $r_0$, respectively, extrapolated
to the present epoch. With a top-hat filtering, the functional form of
$\Delta (r_0)$ for the CDM power spectrum of density perturbations is taken
from Bardeen et al. (1986), and its value is normalized at a scale of
$8h^{-1}$ Mpc with the bias parameter $b_g$.

The number of absorption systems per unit redshift with an H~I column density
larger than $\nh$, that is produced in halos with circular velocity
$V_l < V_c < V_u$, is then given as (Sargent et al. 1980; Mo et al. 1993),
\begin{equation}
\frac{d N(\nh, z, V_l, V_u)}{dz} = \frac{c}{H_0} (1+z)^{1/2} \> \int 
_{V_l} ^{V_u} dV_c \epsilon n(V_c, z) \pi b^2(\nh, V_c) \> , 
\end{equation}
for the EdS universe. We define $(V_l, V_u)= (V_1, V_2)$ for minihalos and
$(V_2, V_3)$ for galactic halos. Here $\epsilon$ denotes the fraction of
halos that give rise to $\lyal$ absorptions, in view of the fact that some
fraction of halos, specifically early-type galaxies such as E, S0, dE and dS0
in the scale of galactic halos, are lacking of interstellar gas. Since these
galaxies experienced the burst of star formation and the subsequent gas loss
via galactic wind for only a short period of $\la 1$ Gyr (see e.g. Yoshii
\& Arimoto 1987), the probability of a sightline to pass the protogalactic,
gas-rich stage of these galaxies may be small.
To evaluate $\epsilon$, we adopt the nominal fraction 69\% of late-type
galaxies among all types derived by Postman and Geller (1984) and assume it
to be constant with $z$ for simplicity. Thus,
\begin{equation}
\epsilon = \left\{
  \begin{array}{ll}
  1    &\qquad\mbox{for mini halos}  \nonumber \\
  0.69 &\qquad\mbox{for galactic halos} \>.
  \end{array}\right.
\end{equation}

Then the above equation can be used to find the fraction of $\lyal$ lines with
$\geq \nh$ that are associated with galactic halos:
\begin{equation}
f_g \equiv \frac{{\rm galactic}}{{\rm mini} + {\rm galactic}} = 
\frac{ d N(\nh, z, V_2, V_3)/dz }{ d N(\nh, z, V_1, V_3)/dz } \>. 
\end{equation}

Besides the case of the `standard' EdS universe, we will also
calculate the number of $\lyal$ lines $dN/dz$ and their galactic fraction
$f_g$ for a world model with a cosmological constant.
The basic cosmological parameters that define the
evolution of the universe are the density parameter $\Omega_0 \equiv
\rho_0 / \rho_c$, where $\rho_c$ is the critical density to close the universe,
and the cosmological constant $\lambda_0 \equiv \Lambda c^2 /3H_0^2$.
We investigate a low-density ($\Omega_0 <1$), flat ($\Omega_0+\lambda_0=1$)
universe, since such a universe deserves special attention to solve several
paradoxical results of cosmological observations (Ostriker \& Steinhardt 1995).
We use the corresponding set of equations (10)-(12) derived from the spherical
collapse theory of density perturbations in a non-zero $\lambda_0$ universe
(see Peebles 1980 and Suto 1993 for details). The model universes which
we adopt for the following analysis are $(\Omega_0,~\lambda_0,~h)=(1,~0,~0.5)$
and $(0.4,~0.6,~0.65)$. Both universes yield the cosmological age as
$\simeq 13$ Gyr, so that any differences in the final results, which may arise
from the different cosmological age, are eliminated.

\section{MODEL RESULTS}
\subsection{{\it The number of absorption lines}}

We first calculate the number of $\lyal$ absorption lines based on the model
described in the last section, and investigate whether or not the model results
can be reconciled with observations.

Figure 2(a) shows the redshift evolution of the number of $\lyal$ absorption
lines with an H~I column density larger than $N_{HI}=10^{14}$ cm$^{-2}$.
The absorption lines arise from both minihalos and galactic halos,
and the bounds on circular velocities are taken as $V_1=15$ km~s$^{-1}$ and
$V_3=250$ km~s$^{-1}$. We take a standard set of parameters for photoionized
gas ($T=3\times10^4$ K for minihalos, $T_{cl}=3\times10^4$ K and
$M_{cl}=10^6 M_{\odot}$ for galactic halos, with $F=0.05$ and $\tj21=1$).
Solid lines correspond to a `standard' EdS universe
($\Omega_0=1$, $\lambda_0=0$), while dotted lines to a low-density
($\Omega_0=0.4$), flat ($\Omega_0+\lambda_0=1$) universe.
It is found, as was already argued by Mo et al. (1993), that
there are two stages in the evolution of the number density. First,
the rapid formation of minihalos inherent in the CDM models gives rise to
the abundant of $\lyal$ lines at $z \ga 10-20$, and second, at lower redshifts
of $z \la 10$, the number of $\lyal$ lines decreases
with time as the hierarchical mergings proceed --- small-sized, less-massive
objects disappear by being incorporated into larger, more massive objects.
The rapid decrease of $dN/dz$ with decreasing $z$ is now an observationally
established fact, first recognized by Peterson (1978).
For the cases of large $b_g$ and non-zero $\lambda_0$, the maximum value
of $dN/dz$ is reduced and the epoch at which it is realized is delayed. This
is because the growth of density perturbations and subsequent mergings are
suppressed due to the bias in galaxy formation and the rapid expansion of the
universe. However, the effects of these different assumptions on model
parameters appear to be almost indistinguishable at the second stage
of the number evolution of $\lyal$ lines. This indicates that probing
world models by $dN/dz$ of absorption systems in the observed redshift
region of $z \la 3$ may not be useful in view of some model uncertainties.
We note here that numerical simulations with a variety of world models
have shown almost similar H~I column density distribution for $\lyal$
lines (Miralda-Escud\`e \etal 1995; Hernquist \etal 1995).
Also plotted in the figure
as filled squares are the observed numbers of $\lyal$ lines at $z \sim 3$
(Petijean et al. 1993), $z \sim 1.5$ (Lu et al. 1991), and $z \sim 0$ (Bahcall
et al. 1993). It is clear that the models with $\j21 \sim 1 = const.$ at
$z \ga 2.5$ are reasonably in agreement with the observed values in the
corresponding range of redshifts, whereas the {\it Hubble Space Telescope}
(HST) result of the unexpectedly large number of low-redshift $\lyal$ lines
(Bahcall et al. 1993), compared to the extrapolated value from high $z$,
can be explained if the UV flux is decreasing with time (Mo et al. 1993).
We find that $\j21 = [(1+z)/3.5]^\alpha$ with $\alpha \sim 2$ fits well.
It is interesting to note that with such a redshift evolution, the current
UV flux is $\j21 (z=0) \sim 8 \times 10^{-2}$, which is consistent with the
observations of Maloney (1992) and Kulkarni and Fall (1993).

The dependence of $dN/dz$ on the column density $N_{HI}$ is shown in
Fig.2(b), where filled squares are taken from Rauch et al. (1992)
for $2 \la z \la 3$. The observed power-law decline of the $\lyal$ lines,
$\propto N_{HI}^{-5/3}$, is well reproduced by the model on the basis of
the isothermal density profile of minihalos (Rees 1986), together with the
current model of pressure-confined clouds in galactic halos. We remark that
this is the case as long as the effect of self-shielding against the external
UV radiation is neglected. The effect, leaving a cooled neutral core in the
inner part of a sphere (e.g. Chiba \& Nath 1994), is realized at the high
column-density ends of $N_{HI} \ga 10^{17-18}$ cm$^{-2}$. This effect is thus
unimportant in the present study of low column-density $\lyal$ lines of
$10^{14} \la N_{HI} \la 10^{16}$ cm$^{-2}$ that occur at the large impact
parameters of sightlines (Fig.1).

\subsection{{\it Fraction of Ly$\alpha$ lines associated with galactic
halos}}

We now investigate what fraction of $\lyal$ lines can be
associated with galactic halos where production of metals is possible --- the
fraction $f_g$ which may be compared to that of $\lyal$ lines associated with
metal lines, $f_m$.

Figure 3(a) shows the evolution of the galactic fraction $f_g$ of $\lyal$
lines with $N_{HI}>10^{14}$ cm$^{-2}$, where we assume $V_2=55$ km~s$^{-1}$ as
the boundary value of $V_c$ between minihalos and galactic halos. It is found
that initially at high redshifts ($z \ga 20$), most of $\lyal$ lines are
provided by minihalos that collapsed early for the CDM-type power spectrum of
density perturbations. This leads to the small values of $f_g$ less than 0.2.
Then, at $10 \la z \la 20$, the fraction of lines in galactic halos is rapidly
increasing with time as a result of the formation of massive halos
via hierarchical merging. Finally at $z \la 10$, the value of $f_g$ slowly
converges to the current fraction of $0.75 \sim 0.80$. The figure also
indicates that the differences arising from the different bias parameters
$b_g$ and cosmological parameters $(\Omega_0,\lambda_0)$, which
are clearly visible at high redshifts, become less significant at lower
redshifts where observations of absorption lines are accessible. All of the
predicted galactic fractions $f_g$ at $z \sim 2.5$ are in between
the reported values of $f_m$ for metal lines, $f_m \sim 0.5 - 0.6$
(Cowie et al. 1995; Tytler et al. 1995; Womble et al. 1996) and $f_m \sim 0.75$
(Songaila and Cowie 1996).

The dependence of $f_g$ on the column density $N_{HI}$ at $z=2.5$
is shown in Fig.3(b). 
There is an indication of the increase of $f_g$ with $N_{HI}$;
it would tend to be 1 at $N_{HI}$ around $10^{20-21}$ cm$^{-2}$
for damped $\lyal$ clouds. There are no observational support yet for
the above behaviour of $f_g$ with $N_{HI}$ and will be a test for the
two population model. Recently, Songaila and Cowie (1996) claim that
$f_m \sim 0.9$ for $N_{HI} > 1.6 \times 10^{15}$ cm$^{-2}$ and
$f_m \sim 0.75$ for $N_{HI} > 3.0 \times 10^{14}$ cm$^{-2}$, whereas the other
observations give a much lower value of $f_m$. More observations are needed
to settle this issue (see also discussion).

As we mentioned in \S 2, the exact value of $V_2$, which separates the
range of circular velocities into those of minihalos and galactic halos,
is yet to be settled from theoretical points of view. For example, even for
$V_c \la 55$ km~s$^{-1}$ where the virial temperature is $T \la 10^5$ K,
some fraction of gas in an inner part of minihalos may be able to cool and
form stars, because of self-shielding effects against the external
UV background radiation. We thus turn the argument and address the question,
as to which range of $V_2$ is allowed in the light of the Keck results.
Figure 4 shows the galactic fraction of $\lyal$ lines as a function of $V_2$,
while $V_1 = 15$ km~s$^{-1}$ and $V_3 = 250$ km~s$^{-1}$. Here we note that
the change of $V_2$ is applied to the range of integration in eq.(12) and the
definition of $\epsilon$ in eq.(13), whereas the range of $V_c$ for the
expressions of $N_{HI}$ ($V_c < 55$ km~s$^{-1}$ for eq.5 and
$V_c \ge 55$ km~s$^{-1}$ for eq.8) is unchanged on the theoretical grounds.
For the observed fraction of $0.5 \sim 0.75$, the possible value of $V_2$
to reproduce it is 40 to 60 km~s$^{-1}$. This is rather
insensitive to the specific assumptions on the bias parameter $b_g$,
cosmological parameters ($\Omega_0,\lambda_0$), and the limiting
circular velocities ($V_1,V_3$). However, to set a more robust
constraint on the range of circular velocities or masses for mini/galactic
halos, further observational determinations of the fraction of $\lyal$
lines associated with metal lines are needed.

\subsection{{\it Metallicity of absorption systems}}

Several recent observations have also measured the metal abundances of
absorption lines at high redshifts (Cowie et al. 1995; Womble et al. 1996).
The presence of heavy elements implies that interstellar gas confined in dark
halos has been processed by star formation and subsequent chemical evolution.
While there are some alternative possibilities for enriching gas --- one
representative idea is that IGM as a whole has already been processed on a
sub-galactic scale {\it prior to} formation of collapsed galactic halos
(Songaila \& Cowie 1996), we present here the hypothesis that enrichment
occurs {\it after} halo formation.
We postulate that gas within halos has been cooled quickly enough to fragment
into stars, and being enriched. In our model, this is applied to the gas in
galactic-sized halos with $V_2 \leq V_c \leq V_3$, where cooling time is well
less than dynamical time (Dekel \& Silk 1986; Efstathiou 1992). Here, we
attempt to estimate the possible metal abundances of gas in these halos.

Suppose that a halo with $V_c$ at the redshift $z$ is enriched with an average
metallicity $Z(V_c,z)$. Here the dependences on $V_c$ and $z$ arise from the
fact that the halos with different $V_c$ collapse and form at different epochs,
say $z_f$, leading to different epochs of star formation, say
$z_{SF}$, and that the subsequent chemical evolution starting at $z_{SF}$
yields the metals at the redshift $z$. Thus an ensemble of these halos with
the metallicity $Z(V_c,z)$ contribute to the metal abundance of $\lyal$ lines.
Then since the probability for a sightline to encounter galactic halos in the
velocity range of $V_2 \leq V_c \leq V_3$ is given by eq.(12), we arrive at the
average metallicity of $\lyal$ lines with H~I column density larger than
$N_{HI}$:
\begin{equation}
<Z> (>N_{HI},z) =
 \frac{\int_{V_2}^{V_3} dV_c \cdot Z(V_c,z) n(V_c, z) \pi b^2(\nh, V_c)}{
 \int_{V_2}^{V_3} dV_c \cdot n(V_c, z) \pi b^2(\nh, V_c)} \> . 
\end{equation}

The actual expression for $Z(V_c,z)$ in each halo is admittedly not
straightforward to derive and very model-dependent. Physics 
involved in it
are, e.g. merging history of a population of dark halos, gasdynamical and
thermal processes of gas confined in these halos, and history of star
formation. While the extensive and sophisticated approaches of incorporating
these processes exist (Lacey \& Cole 1993; Kauffmann et al. 1993), we adopt a
more simplified approach which we believe includes the essential properties
of the system to deduce the characteristic values of metal abundance.
First, we assume that the formation redshift $z_f$ of the halo with mass
$M$ is represented by the collapse epoch of a rms, 1~$\sigma$ linear density
contrast, i.e., if $D(z)$ denotes the growth rate of density perturbation,
$z_f$ is derived from $D(0)/D(z_f) = \Delta(M)/\delta_c$, where $M$
is related to $V_c$ from eq.(10). According to Lacey \& Cole (1993)
on the basis of their extensive analysis of merging
halos, this scaling relation appears to hold for the typical formation
epoch of the {\it main parent} halo which had half or more of the present
mass $M$ (see their Fig.9 and Fig.10). Realistically the halo formation epoch
for a given $M$ is distributed with some probability arising from different
amplitudes of density perturbations, but it can be characterised by the
collapse epoch of a typical, 1~$\sigma$ perturbation. Second, gas in a
virialized, galactic-sized halo must cool quickly (within  a
dynamical time) to ensure star formation, so that $z_{SF}$, the epoch of star
formation, is assumed to be equal to $z_f$. This is reasonable, because
the cooling time in the galactic scales is less than $10^8$ yr, which is
a negligible time span --- much less than the Hubble time --- in the
cosmological context. Third, concerning the subsequent chemical evolution,
we adopt the observed age-metallicity relation of metal-poor halo stars in our
Galaxy (Schuster \& Nissen 1989), on the grounds that typical sightlines to
produce the $\lyal$ lines with low H~I column density penetrate the halo gas,
from which old halo stars may have formed. The average enrichment rate
$\Delta Z / \Delta t$ derived by Schuster \& Nissen (1989) is of the order of
$10^{-3}$ Gyr$^{-1}$ for their sample of halo stars. This value is also close
to the value from the chemical-evolution model of the halo by Pagel (1989).
Therefore, denoting the age of the universe
at the redshift $z$ is $t_{age}(z)$, the metallicity $Z$ at $z$ is derived as,
\begin{equation}
Z(V_c,z) = \int_{t_{age}(z_{SF}(V_c,z))}^{t_{age}(z)}
 \frac{\Delta Z}{\Delta t} dt
 \> , 
\end{equation}
where following the above arguments, the epoch of star formation,
$t_{SF} \equiv t_{age}(z_{SF}(V_c,z))$, depends on
$V_c$ and $z$.

We assume here that at the outer parts of the halo, which concerns us
here for the $\lyal$ forest lines, the gradient of metallcitiy is negligible.
This is a reasonable assumption because there are indications that
star formation in the outer halo of our Galaxy has been slow and,
therefore, the chemical evolution (Matteucci and Fran\c{c}ois 1992).
If the variation in
star formation rate is caused by the variation in the gas collapse time,
then this variation is expected to be small in the outer regions of the halo.
This assumption is consistent with the observation that the average metallicity
is almost constant for $\lyal$ lines with $N_{HI} \la 10^{16}$ cm$^{-2}$
(Cowie \etal 1995, Songaila and Cowie 1996).

Figure 5(a) shows the evolution of the mean abundance for absorption lines
with $N_{HI} > 10^{14}$ cm$^{-2}$ (solid and dotted lines). It is found that
for $3 \la z \la 5$, the metal abundance significantly increases by more
than two orders of magnitude. At $z \sim 2.5$, it is approximately
in agreement with the Keck results of [C/H]$=-2.5 \sim -2$.
We remark that the adopted age-metallicity relation of a constant enrichment
rate overestimates the halo abundance at $z \la 1$: the relation should
be applied to
stars with metallicity less than $10^{-1}$ of solar, which is
the criterion to separate halos stars from disk stars (Schuster \& Nissen
1989). The subsequent formation of halo stars is stopped, and most of the
enriched halo gas would fall into the disk --- the hypothesis supported
for solving the so-called G-dwarf problem in the solar neighborhood (e.g.
Yoshii et al. 1996). Thus we assume $\log <Z>/Z_{\odot} = -1$ as an upper
limit for abundance of absorption lines in Fig.5(a), but this does not change
our result in the light of the Keck results at $z = 2 \sim 3$.
We have also investigated the case when the epoch of star
formation $t_{SF}$ is constant, being independent of the mass scale of halos
--- as was done by Timmes et al. (1995) for damped $\lyal$ systems.
We find that the evolution of $<Z>$  based on our prescription of
star formation beginning at the collapse epoch of 1 $\sigma$
density contrasts is essentially the same as the case
when $t_{SF} = 1.5 \sim 2$ Gyr. 

Also shown in Fig.5(a) with filled circles are the recent findings of
Lu, Sargent \& Barlow (1996) for the abundances ([Fe/H])
of damped $\lyal$ systems observed with the Keck. Intriguingly, the median
values of abundances in such systems are well reproduced by the current models.
This may support the hypothesis, first proposed by Lanzetta, Wolfe \& Turnshek
(1995a), that damped $\lyal$ systems could represent a spheroidal component
of galaxies (see discussion in \S4.2).

The dependence of $<Z>$ on the value of $V_2$ is shown in Fig.5(b). In contrast
with the galactic fraction $f_g$ of $\lyal$ lines, $<Z>$ turns out to be
sensitive to both the bias parameter $b_g$ and the cosmological parameters
($\Omega_0,\lambda_0$) at large values of $V_2$. It is systematically
lower for a larger bias and/or lower-density universe. This is caused by
delayed star-formation epoch $t_{SF}$, leading to less time for enriching gas.

\section{DISCUSSION AND CONCLUDING REMARKS}
\subsection{{\it Evolution of galactic fraction and average metallicity}}

We have shown that the observed fraction of $\lyal$ absorption lines associated
with metal lines can be understood in terms of photoionized gas confined in
galactic-sized halos with $40-60 \la V_c \la 250$ km~s$^{-1}$, whose number
density follows the Press-Schechter model of hierarchical structure formation.
The derived average abundance of heavy elements in absorption systems, combined
with the characteristic epochs of star formation and the observed
age-metallicity relation in our Galactic halo, appears to be reasonably in
agreement with the observed abundance at $z=2 \sim 3$.

The current model also suggests that both the fraction of $\lyal$ lines
associated with galactic halos and the metallicity in absorption systems
are expected to increase with decreasing redshift (Fig.3a and Fig.5a). While
the explicit dependences of these quantities on redshift are yet to be
settled (see Songaila \& Cowie 1996), we note the findings of
Steidel (1990) that the number density of C~IV lines increases with
decreasing redshift. In our model, this can be explained from the above
predicted properties of absorption systems in the following way.

In photoionization equilibrium, the ratio between the column densities
of C~IV and H~I is a function of the ionization
parameter $\Gamma$, which is the ratio of the densities of ionizing
photons and particles, and of the metallicity $Z$; using the photoionization
code of CLOUDY (Ferland 1993), we find that for $\log_{10} \Gamma =-1.5$
to $-2.0$, and for $Z=0.001$ to 0.5, the ratio $N_{CIV}/N_{HI}$ is well
described as,
\begin{equation}
\frac{N_{CIV}}{N_{HI}} \propto \Gamma ^{-1.3 \pm 0.2}  Z ^{1.0 \pm 0.1}
 \> . 
\end{equation}
Then, suppose that the number density of $\lyal$ lines per unit H~I column
density and redshift, $d^2\bar{N}_{HI}/dN_{HI}dz$, is represented by a power
law form,
\begin{equation}
\frac{d^2\bar{N}_{HI}}{dN_{HI}dz} d N_{HI} dz \propto
   N_{HI} ^{- \beta} (1+z)^ {\gamma} d N_{HI} dz \> , 
\end{equation}
where $\beta \sim 1.5$ (Sargent et al. 1989) and $\gamma \sim 2.45$
(Lu et al. 1991), and that a fraction $f_g$ of these $\lyal$ lines are
associated with C~IV lines, which may be written as,
\begin{equation}
f_g(z) \propto (1+z)^{- \eta}  \> .  
\end{equation}
From Fig. 3(a), we obtain $\eta \sim 1.6$ at $z = 2 \sim 3$.

Suppose also that the mean metallicity of absorption lines increases in time
following $<Z> \propto (1+z) ^{- \delta}$. Then eq.(17) and eq.(18) give the
column density distribution of C~IV lines as,
\begin{equation}
\frac{d^2\bar{N}_{CIV}}{dN_{CIV}dz} d N_{CIV} dz \propto
   N_{CIV}^{- \beta} (1+z)^{\gamma_C} d N_{CIV} dz \> , 
\end{equation}
where
\begin{equation}
\gamma_C= \gamma  - \delta (\beta)  - \eta \> , 
\end{equation}
describes the redshift evolution of C~IV lines.
It is known from observations that the $\beta$ index is the same 
for both H~I and C~IV
lines (Cowie et al. 1995; Bergeron et al. 1994),
indicating that the simple equation (17) holds well.

Steidel (1990) has shown that the index $\gamma _C$ is negative, of
order $-1.25$. This means that the number of C~IV lines increases in time.
In our model, this can be due partly to the increase of the galactic fraction
$f_g$ in time and to the increase of the metallicity $<Z>$ in time.
Then, for the set of power-law indices, $\beta =1.5$, $\gamma _C=-1.25$,
$\gamma=2.45$, and $\eta=1.6$, we arrive at $\delta =1.4$, thereby indicating,
\begin{equation}
<Z> \propto (1+z)^{-1.4} \> . 
\end{equation}
It is worth noting that the enrichment rate of absorption lines shown
in Fig.5(a) is well approximated by this equation.
The redshift evolution for C~IV lines used here is, strictly speaking,
for high C~IV column density lines. It will be interesting to see
whether observations in the near future show a similar redshift
evolution for small C~IV column density lines.

Therefore, the derived redshift evolutions of both the fraction of $\lyal$
lines associated with galactic halos (Fig.3a) and the metallicity in absorption
systems (Fig.5a) are approximately consistent with the observed  scaling
laws for $\lyal$ and C~IV lines.

\subsection{{\it Possible relation with metals in damped Ly$\alpha$
systems}}

The abundances of metals have also been measured in damped $\lyal$
systems, which are believed to be protogalaxies. Smith et al. (1996)
and Pettini et al. (1995) have recently reported, by measuring the Zn lines as
a probe of metallicity, that the typical Zn abundance at $z \sim 2$ is $1/15$
of solar, but there is a considerable scatter in the Zn abundance,
spanning more than two orders of magnitude. Using high quality spectra
obtained with the Keck telescope, Lu, Sargent \& Barlow (1996)
have also derived a large scatter of abundances, about a factor of 30,
with $N_{HI}$-weighted mean abundance of 0.028 solar between $2 < z < 3$.
In the context of the present model, such a large scatter may be caused by
the different formation epoch of halos, $z_f$, for a given mass arising from
the different amplitudes of primordial density fluctuations and different
merging history (Lacey \& Cole 1993).
Also, for the adopted age-metallicity relationship of halo stars,
there is a scatter in the ages of the stars of $\pm 2.5$ Gyr at a given
abundance, leading to a scatter of abundances more than an order of magnitude
at a given age (Schuster \& Nissen 1989). It is worth noting that
the evolution of the mean abundance is reproduced
from the formation epoch of a typical 1 $\sigma$ fluctuation and the mean
age-metallicity relationship of halo stars in our Galaxy (Fig.5a).
This suggests that damped $\lyal$ systems represent a metal-poor halo
component. Also, Songaila and Cowie (1996)
found that the ratio Si/C is about three times solar, similar to the
abundance ratio of Galactic halo stars.

\subsection{{\it The possibility of an enriched IGM}}

The two population model, as presented here, clearly predicts (a) the
fraction of $\lyal$ lines with associated metal lines as a function of 
redshift and (b) the redshift evolution of the
average metallicity. The observations, however, have not yet converged to any
firm conclusion and it is too early to rule out other possibilities.
As noted in Songaila and Cowie (1996), it will be very difficult to
determine the fraction of metal enriched $\lyal$ clouds below
$N_{HI} \la 10^{15}$ cm$^{-2}$, because this will mean a sensitivity
limit of $N_{CIV} \la 10^{11}$ cm$^{-2}$, which is difficult at
present. They claim to have found associated C~IV lines in $90 \%$ of clouds
with $N_{HI} > 1.6 \times 10^{15}$ cm$^{-2}$ and $75 \%$ of clouds
with $N_{HI} > 3.0 \times 10^{14}$ cm$^{-2}$. If confirmed,
such a large fraction of enriched clouds will be difficult to
explain with the two population model. As emphasized by
Songaila and Cowie (1996), the alternative hypothesis of an enriched
IGM will then be more appropriate.

We briefly note here that a
problem also prevails at low redshifts about the interpretation of the
fraction of $\lyal$ lines with associated metal lines. Although the association
of some of the low column density $\lyal$ lines with galaxies
seems sure, it is still uncertain as to how physical the association is
(see Le Brun \etal 1996 for a discussion). It is possible that instead
of being directly associated with galactic halos (as in, for example,
Mo and Morris 1994), the $\lyal$ lines
trace some structure related to formation of galaxies (Morris and
van den Bergh 1994,
Petitjean \etal 1995).

We note here the possibility, as shown in several recent numerical
simulations (Miralda-Escud\`e \etal 1995, Hernquist \etal 1995) that
very low column density $\lyal$ lines may arise in mini pancakes and
not minihalos. In this case, the density profile will be much different
and our results may not hold. However, we note that mini pancakes can
account for $\lyal$ lines with $N_{HI} \la 10^{15}$ cm$^{-2}$ (Miralda-Escud\`e
\etal 1995) and
lines with higher H~I column density come from the intersecting points
of pancakes, and correspond to minihalos or galactic halos of our model.
Therefore, our results are most certain to hold for $N_{HI} \ga 10^{15}$
cm$^{-2}$. In any case, as explained above, the  fraction of $\lyal$
lines with associated metal lines for $N_{HI} \la 10^{15}$ cm$^{-2}$
will be difficult to determine in the near future, and our model will
not be testable for such values of H~I column densities.

Incidentally, the case for an enriched IGM was described by Silk
\etal (1987) much before the discovery of associated metal lines in
$\lyal$ systems. They discussed the possibility of enriching the IGM
by galactic winds from dwarf galaxies. The detail of such a scenario,
however, remains to be worked out. Pregalactic objects such as Population
III stars might also enrich the IGM at high redshifts, but again
no detail models exist.

In the face of such interesting but still poorly understood scenarios,
it therefore seems reasonable to pursue a modified version of the minihalo
model, such as the two population model presented here. We have
shown that a fraction of $\ga 0.5$ of $\lyal$ forest lines
with associated metal lines can be understood in the framework of
hierarchical structure formation model, assuming that halos with
velocity dispersion $15 \la V_c \la 55$ km/s do not produce stars
and heavy elements. The model predicts that the fraction of clouds with
associated metal lines increases slowly with the H~I column density.
We have shown that the results are robust and
do not depend strongly on the assumptions of the limiting values of
velocity dispersion. We have further used the age-metallicity relation
of Galactic halo stars to predict the evolution of the average metallicity
of the associated metal line systems. In particular, we find that the average
metallicity at $z \sim 2$ should be about a hundredth of solar.

\acknowledgments

We thank T.~Padmanabhan, R.~Srianand, and E.~J.~Wampler for stimulating
discussions and helpful comments. We are also grateful to the referee for
his constructive suggestions for the improvement of this paper.
This work has been supported in part by the Grand-in-Aid for Scientific
Research (08640318) of the Ministry of Education, Science, and Culture
in Japan.

\appendix
\section{$N_{HI}$ vs. $b$ in galactic halos}

We describe below how to obtain the H~I column density as a function of
the impact parameter $b$ for $\lyal$-absorbing gas clouds in galactic halos.

The probability $Pdl$ that a sightline in the range of $l \sim l+dl$
encounters the clouds confined in the ambient hot gas is given as
$Pdl = n_s \sigma dl$, where $n_s$ is the number density of clouds and
$\sigma \equiv \pi R_{cl}^2$ is the cross section of each cloud.
Under the assumption that each sightline encounters one cloud
in average, $P$ is normalized as $\int P dl = 1$. Then, if each cloud has
its own column density $N_{HI}(r)$ as a function of the galactocentric
distance $r$, the average column density $<N_{HI}>(b)$ at the impact parameter
$b$ may be written as,
\begin{eqnarray}
<N_{HI}>(b) &=& \int_0^\infty N_{HI}(r)n_s \sigma dl \nonumber \\
       &=& 2 \int_b^\infty N_{HI}(r) \frac{n_s \sigma rdr}{\sqrt{r^2-b^2}} \ .
\end{eqnarray}

To obtain the possible expression for $n_s$, we assume that the whole area
of a galactic halo projected onto the sky is in average covered by ensemble of
clouds without overlapping. This is equivalent to $\int P dl = 1$ as described
above. Then, the average number density of clouds within $b$, denoted
as $\bar{n}_s$, may be approximately given as
$\bar{n}_s \sim 1/(4\pi R_{cl}^2 b/3)$, and this yields the number density
$n_s$ at each $b$ as $n_s \sim 1/(2\pi R_{cl}^2 b)$.
Equation (A1) is then reduced to 
\begin{equation}
<N_{HI}>(b) = \frac{1}{b} 
       \int_b^\infty N_{HI}(r) \frac{rdr}{\sqrt{r^2-b^2}}
\end{equation}

Here we note that $N_{HI}(r) \propto r^{-10/3} \equiv Kr^{-10/3}$ at
$r \gg r_c$ where H~I column densities of $\lyal$ absorption lines are lower
than $10^{16}$ cm$^{-2}$ in galactic halos. Substituting this into eq.(A2),
we obtain
\begin{equation}
<N_{HI}>(b) = \frac{K}{b} \int_b^\infty \frac{r^{-7/3}dr}{\sqrt{r^2-b^2}}
     = K b^{-10/3} \int_0^\infty \frac{dx}{\cosh^{7/3}x} \ ,
\end{equation}
where we introduce the variable $x$ by $r = b \cosh x$ in the second equation.
The integration in eq.(A3) yields 0.91, and this can be regarded as $\sim 1$
in view of the other uncertainties. This allows us to adopt the approximate
equality as $<N_{HI}>(b) \sim N_{HI}(r \to b)$ .

\clearpage

\clearpage
\noindent{\bf Figure Captions}
\medskip

\figcaption[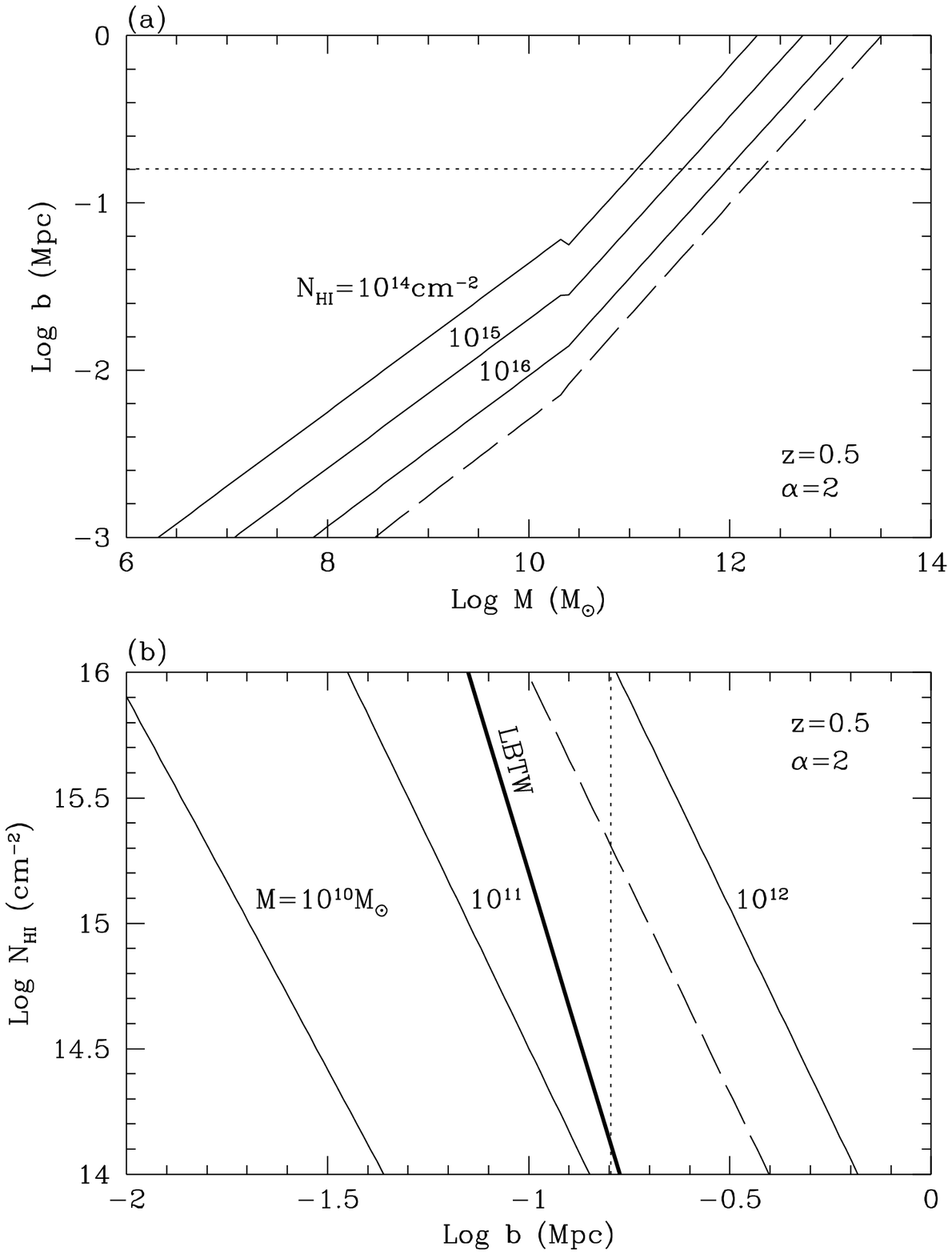,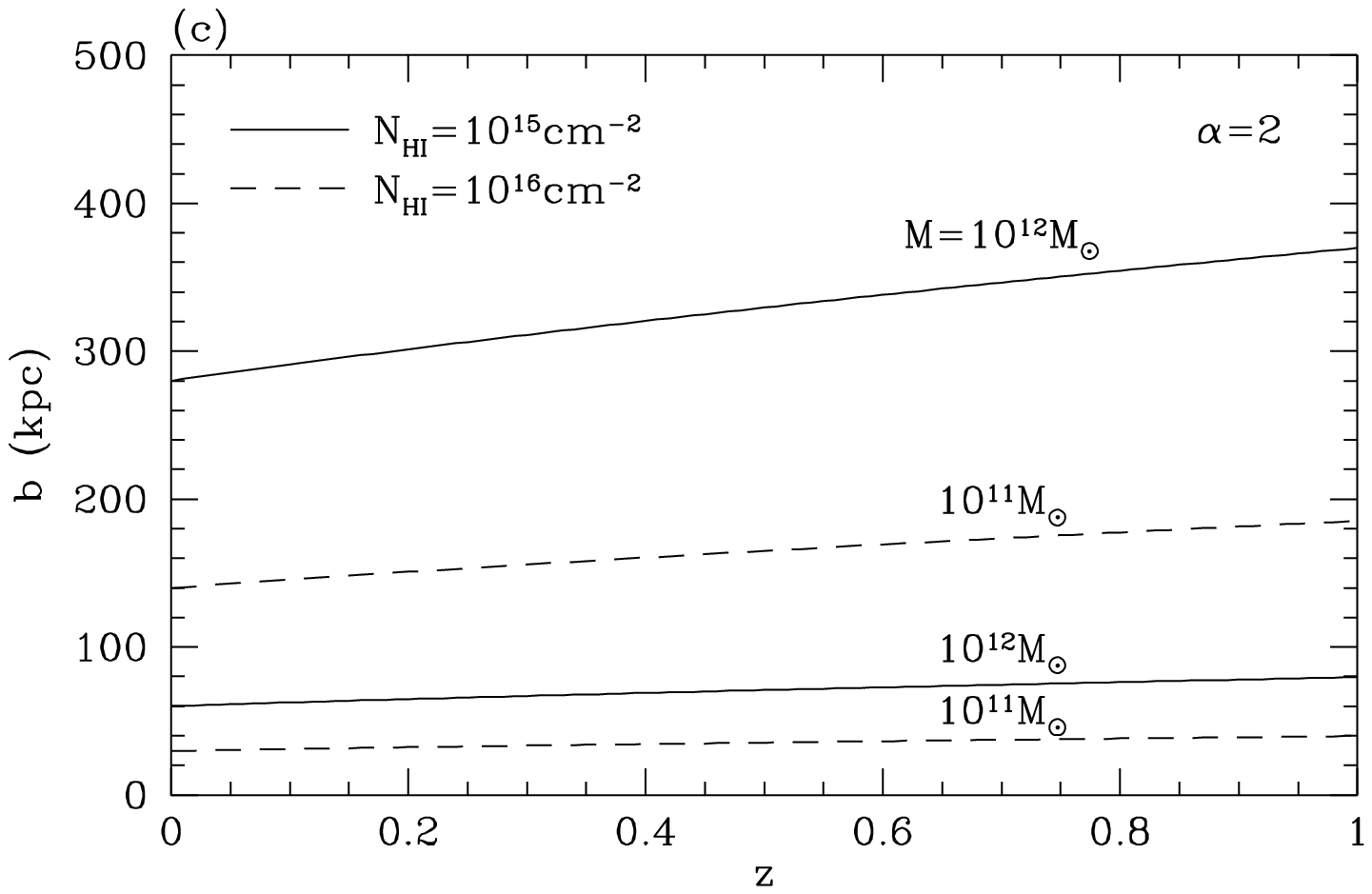]{{\bf (a)} The impact parameter $b$ (Mpc) of a
sightline as a function of the total mass of a halo $M$ ($M_{\odot}$) to
produce an H~I column density $N_{HI}$ at $z=0.5$ ($F=0.05$, $\tj21=1$,
$x_v=15$, and $h=1$). The solid lines correspond to $\alpha=2$ (or
$\j21 \simeq 0.18$ in eq.9) for $N_{HI}=10^{14}$, $10^{15}$, and $10^{16}$
cm$^{-2}$ from left to right, respectively.
The dashed line corresponds to $\alpha=0$ (or $\j21 = 1$) for
$N_{HI}=10^{16}$ cm$^{-2}$. These lines are quite insensitive
to the adopted cosmological parameters of $\Omega_0$ and $\lambda_0$.
For comparison, the observed value of $\sim 160$ $h^{-1}$ kpc for the size of
gaseous halos at $z \la 1$ is plotted as a dotted line (LBTW).
{\bf (b)} The H I column density $N_{HI}$ (cm$^{-2}$) against the impact
parameter $b$ (Mpc). The solid lines correspond to $\alpha=2$ for $M=10^{10}$,
$10^{11}$, and $10^{12}$ $M_{\odot}$ from left to right, respectively, whereas
the dashed line to $\alpha=0$ for $M=10^{12}$ $M_{\odot}$. The other parameters
are the same as those in (a). The dotted line is the observed size of gaseous
halo $\sim 160$ $h^{-1}$ kpc, whereas the bold solid line denotes the reported
correlation between $N_{HI}$ and $b$ (LBTW).
{\bf (c)} The impact parameter $b$ (kpc) of a sightline as a function of
redshift $z$.}

\figcaption[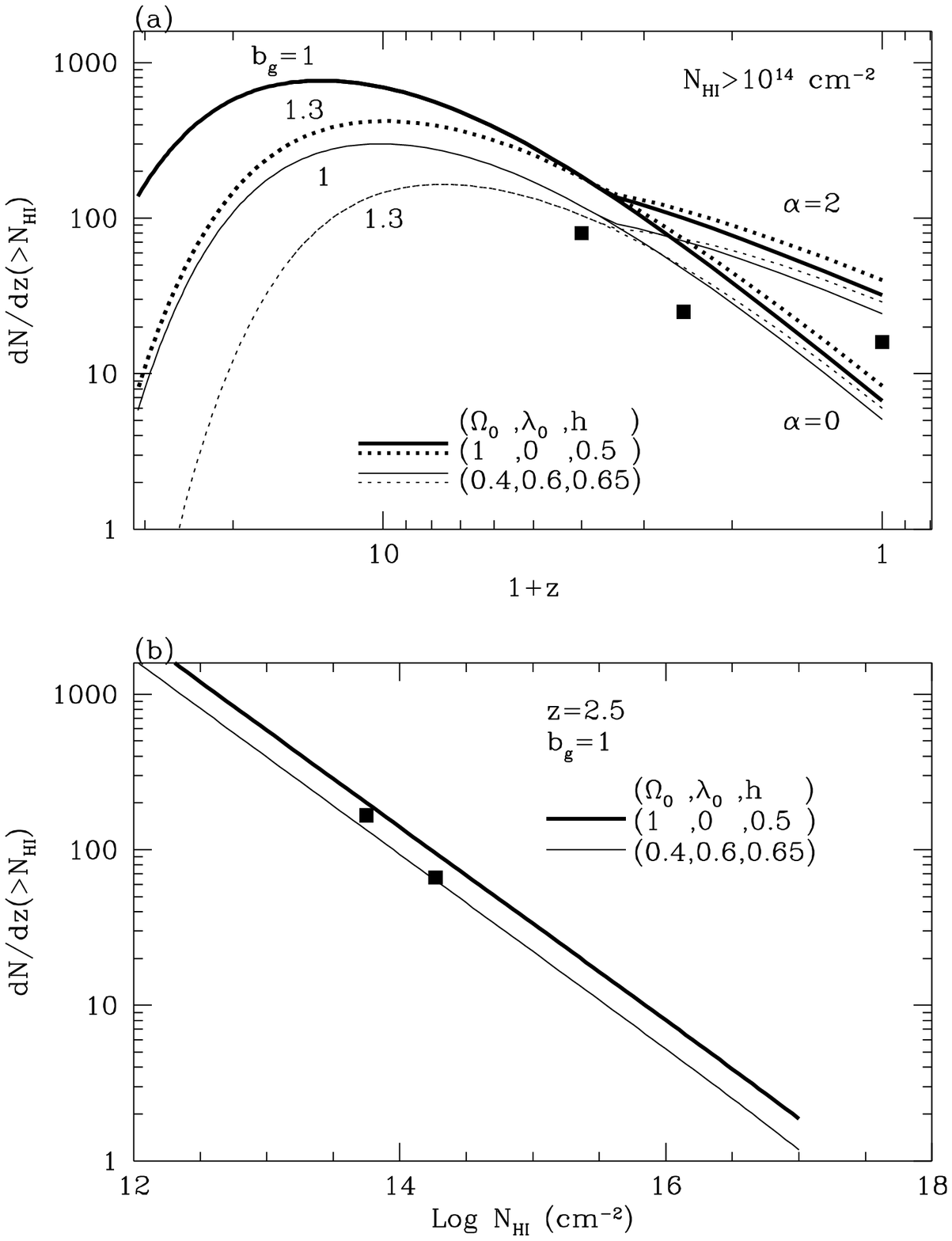]{{\bf (a)} The redshift evolution of the number
of $\lyal$ absorption lines with a column density larger than
$N_{HI}=10^{14}$ cm$^{-2}$ ($F=0.05$, $\tj21=1$, $x_v=15$).
The range of circular velocities $V_c$
are $15 \leq V_c \leq 250$ km~s$^{-1}$ (i.e. $V_1=15$ km~s$^{-1}$,
$V_3=250$ km~s$^{-1}$). Thick and thin lines are for the EdS universe
($\Omega_0=1$, $\lambda_0=0$, $h=0.5$) and the non-zero $\lambda_0$ universe
($\Omega_0=0.4$, $\lambda_0=0.6$, $h=0.65$), respectively, for CDM models with
bias parameter $b_g=1$ (solid lines) and $b_g=1.3$ (dotted). The different
curves at $z \leq 2.5$ corresponds to the different values of $\alpha$ in
eq.(6): the upper curves for $\alpha=2$ and the lower for $\alpha=0$.
Filled squares denote the observed number of $\lyal$ lines at
$z \sim 3$ (Petijean et al. 1993), $z \sim 1.5$ (Lu et al. 1991), and
$z \sim 0$ (Bahcall et al. 1993). {\bf (b)} The dependence of $dN/dz$ on the
H~I column density limit for $z=2.5$ and $b_g=1$. Filled squares denote the
results of Rauch et al. (1992) for $2 \la z \la 3$. The other parameters are
the same as those in (a).}

\figcaption[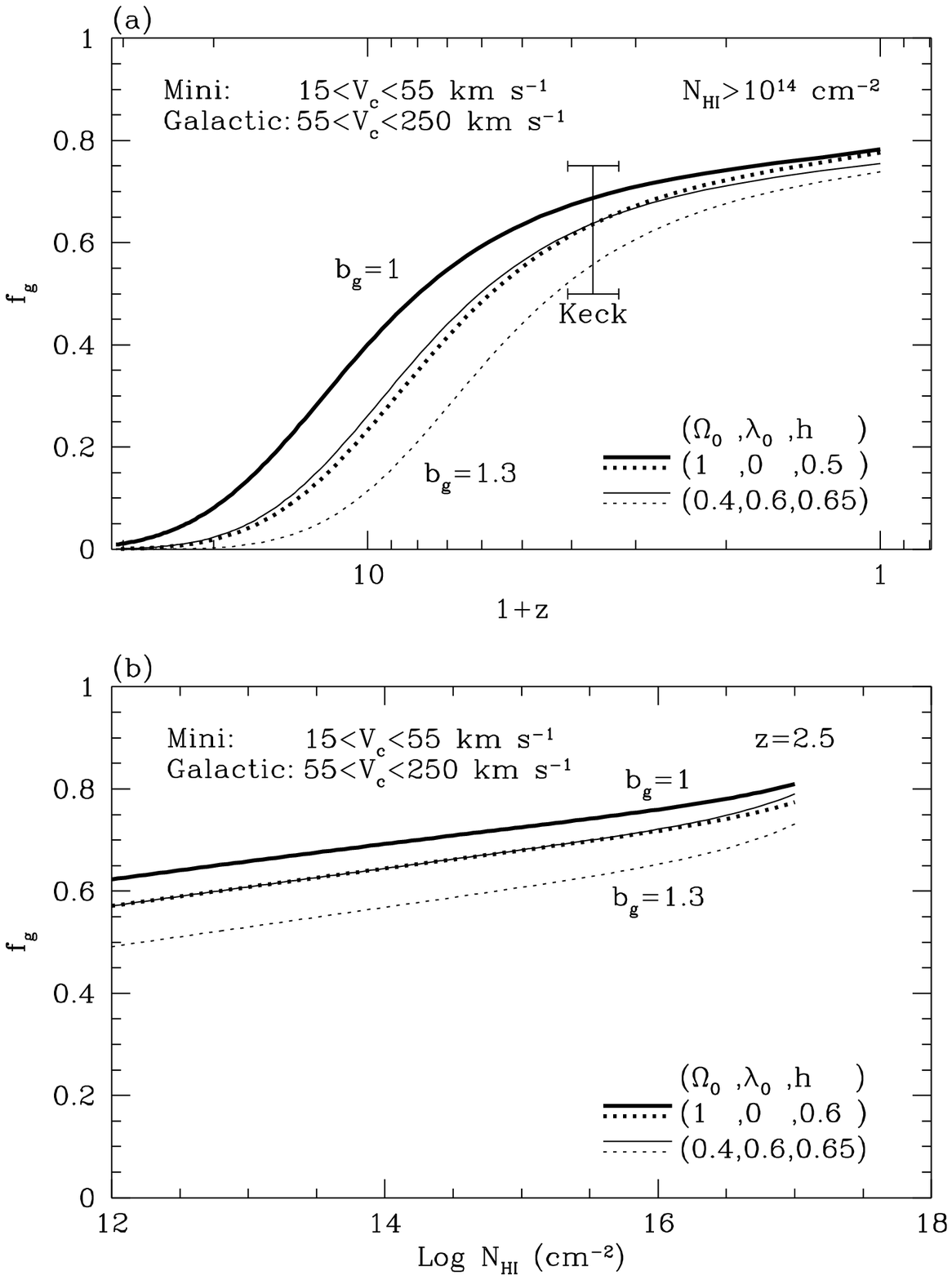]{{\bf (a)} The redshift evolution of the fraction
of $\lyal$ absorption lines associated with galactic halos in the range of
$55 \leq V_c \leq 250$ km~s$^{-1}$ (while $15 \leq V_c \leq 55$ km~s$^{-1}$
for minihalos), with $N_{HI} > 10^{14}$ cm$^{-2}$. The error bar
denotes the Keck results for the fraction of $\lyal$ lines associated with
CIV metal lines (Cowie et al. 1995; Tytler et al. 1995; Womble et al. 1996;
Songaila \& Cowie 1996).
The meanings of the four curves presented are the same as those in Fig.2(a).
{\bf (b)} The dependence of the fraction of $\lyal$ absorption lines associated
with galactic halos on the H~I column density limit at $z=2.5$. The others are
the same for (a).}

\figcaption[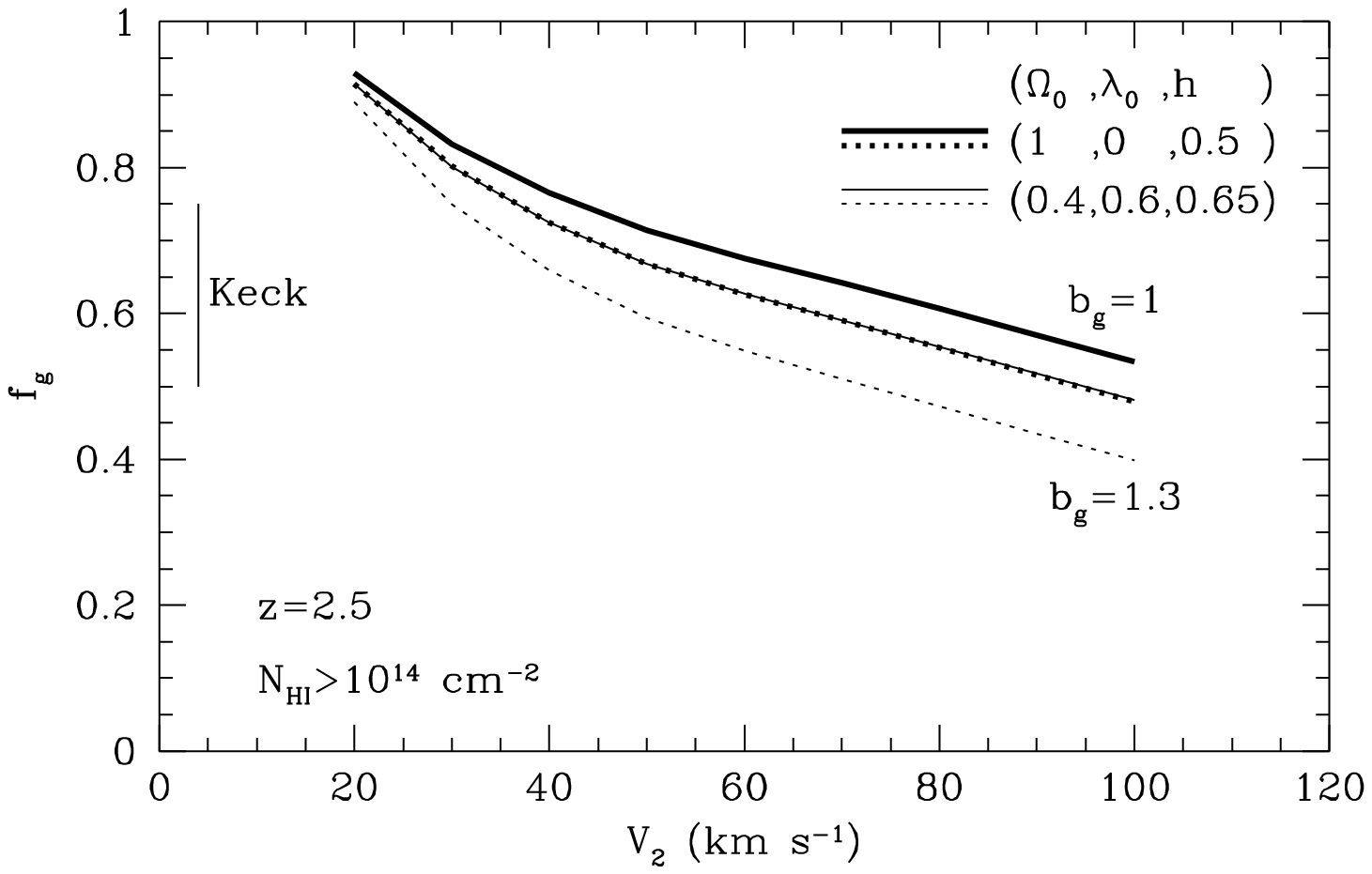]{The dependence of the fraction of $\lyal$ absorption
lines associated with galactic halos on the lower bound of the circular
velocity $V_2$ (while $V_1=15$ km~s$^{-1}$ and $V_3=250$ km~s$^{-1}$),
for $z=2.5$ and $N_{HI}>10^{14}$ cm$^{-2}$. The meanings of the four curves are
the same as those in Fig.2(a). The observed fraction of $\lyal$ lines
associated with CIV metal lines, $0.5 \sim 0.75$, is indicated as a bar.}

\figcaption[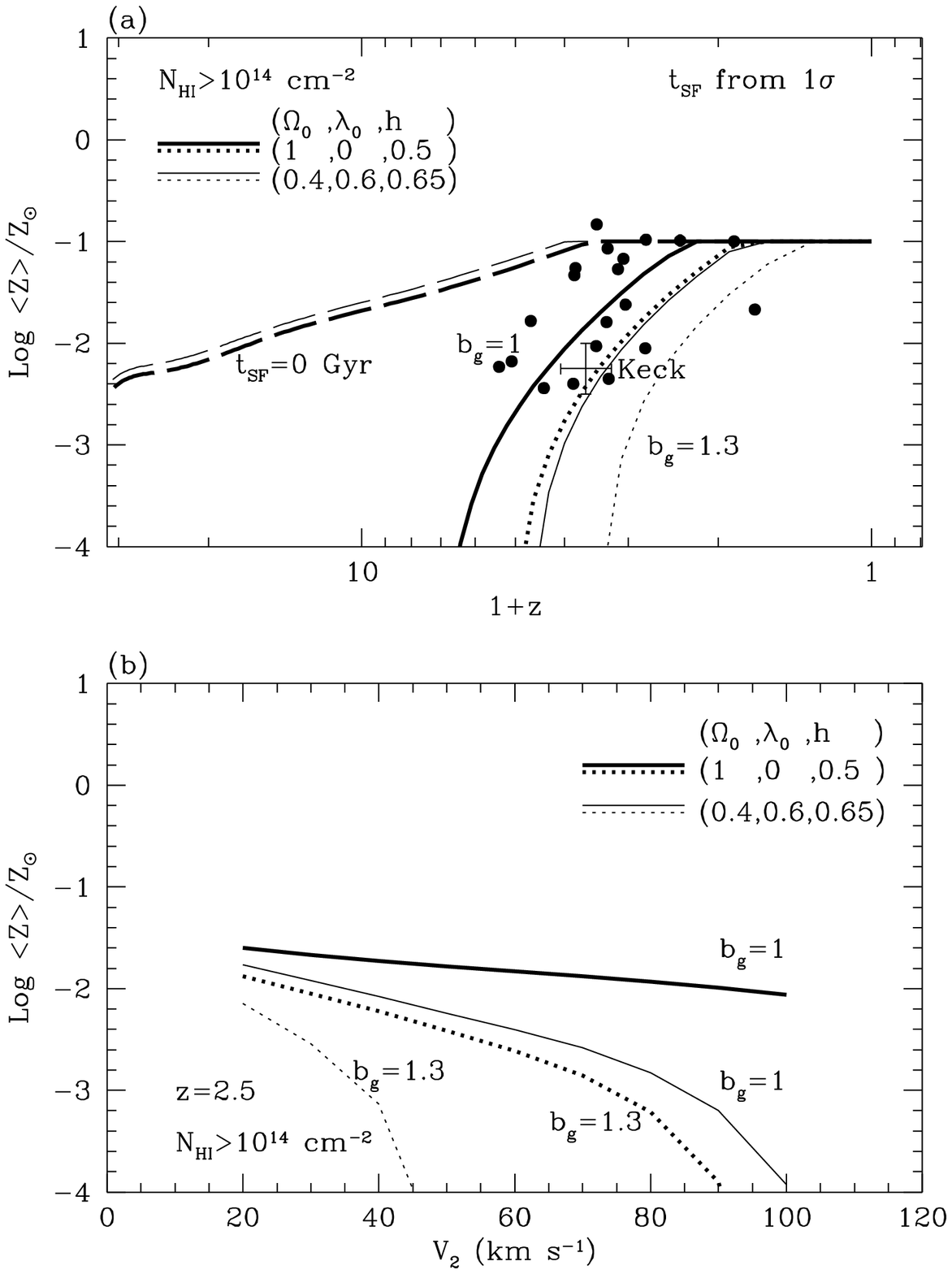]{{\bf (a)} The redshift evolution of the average
metallicity $<Z>$ of absorption lines (solid curves for $b_g=1$ and dotted
for $b_g=1.3$), based on the model described in subsection 3.3. The range of
circular velocities for galactic halos are $55 \leq V_c \leq 250$ km~s$^{-1}$.
For reference, the metallicity evolutions if the star formation commences from
the beginning of the universe, $t_{SF}=0$, are shown as dashed lines.
The crossed error bars denote the Keck results for the $\lyal$ lines
associated with CIV metal lines ([C/H])
(Cowie et al. 1995; Tytler et al. 1995; Womble et al. 1996),
and filled circles denote the abundances
of damped $\lyal$ systems ([Fe/H]) observed with the Keck
(Lu, Sargent \& Barlow 1996).
{\bf (b)} The dependence of the average metallicity of absorption lines
on the lower bound of the circular velocity $V_2$
(while $V_3=250$ km~s$^{-1}$), at $z=2.5$.}

\clearpage
\plotone{fig1ab.eps}
\clearpage
\plotone{fig1c.eps}
\clearpage
\plotone{fig2.eps}
\clearpage
\plotone{fig3.eps}
\clearpage
\plotone{fig4.eps}
\clearpage
\plotone{fig5.eps}
\end{document}